\documentclass[reprint,showpacs,nobibnotes,amsmath,amssymb,aps, pra]{revtex4-2}

\usepackage{graphicx}
\usepackage{dcolumn}
\usepackage{bm}
\usepackage{feyn}
\usepackage{subfigure}
\usepackage{natbib}
\usepackage{float}
\usepackage{color}
\usepackage{amsmath}
\usepackage{amssymb}
\makeatletter

\newcommand{\Rmnum}[1]{\expandafter\@slowromancap\romannumeral #1@}
\makeatother
\usepackage{hyperref}
\hypersetup{hypertex=true,
	colorlinks=true,
	linkcolor=blue,
	citecolor=blue,
	anchorcolor=blue,
	urlcolor=blue}
\usepackage{mathrsfs}
\usepackage{rotating}
\newcommand{\RNum}[1]{\uppercase\expandafter{\romannumeral #1\relax}}
\newcommand{\be}{\begin{eqnarray}}
	\newcommand{\ee}{\end{eqnarray}}

\begin{document}
	
	
\title{Engineering of energy band and its impact on light transmission in non-reciprocal Hermitian hourglass lattice}
	
\author{Junhao Yang}
\affiliation{School of Physics, Northwest University, 710127, Xi'an, China}
\author{Yuandan Wang}
\affiliation{School of Physics, Northwest University, 710127, Xi'an, China}
\author{Yu Lin}
\affiliation{School of Physics, Northwest University, 710127, Xi'an, China}
\author{Wenjing Zhang}
\affiliation{School of Physics, Northwest University, 710127, Xi'an, China}
\author{Guoguo Xin}
\affiliation{School of Physics, Northwest University, 710127, Xi'an, China}
\author{Xinyuan Qi}
\email{ qixycn@nwu.edu.cn}
\affiliation{School of Physics, Northwest University, 710127, Xi'an, China}


\begin{abstract}
	We study a quasi-one-dimensional non-reciprocal Hermitian hourglass photonic lattice that can accomplish multiple functions. Under the effect of non-reciprocal coupling, this lattice can produce an energy isolation effect, two kinds of flat bands, and energy band inversion. The excitation and propagation of a single energy band and multiple energy bands can be realized; in the flat band condition, the system has compact localized states, and the flat bands can be excited by a straightforward method. In addition, we investigate the edge states under the open boundary condition; a double edge state appears by using a defect in the system. Our findings advance the theory of energy band regulation in artificial photonic lattices.
\end{abstract}


\maketitle


\section{Introduction}
Discrete transmission of optical waves, as one of the essential means of spatial optical modulation, is mainly realized by the joint action of trapping in weak waveguides and evanescent coupling between near-adjacent waveguides~\cite{LEDERER20081}. In recent years, most discrete optical transmission research uses periodic photonic lattices as the research platform. Photonic lattice, a renowned physical model with periodically modulated refractive index change in space, is widely used to explore the light dynamics in discrete optical systems, one of the essential platforms for various optical research~\cite{Fleischer2003, PhysRevA.84.021806, RevModPhys.91.015006, LAN2022100076, Price_2022}. In such lattices, the coupling between the adjacent waveguides is symmetrical, and the energy transfer is reversible. Thus, these lattices are called reciprocal coupling photonic lattices, and the same coupling coefficient between the lattice sites characterizes their structures. Furthermore,  some photonic phenomena can be observed through evanescent couplings, such as discrete diffraction~\cite{Pertsch:04, PhysRevLett.83.2726, Zannotti:19}, optical Bloch oscillations~\cite{ PhysRevLett.83.4752, Zhao:22}, dynamic localization~\cite{PhysRevB.34.3625, DENG2022128365}, Zener tunneling~\cite{Fratalocchi:06,PhysRevE.95.042204}. In reality, the shape of a single waveguide can be asymmetric about its transmission axis. That is due to fabrication process limitations or for some particular purposes. As a result, the coupling effect between the near-adjacent waveguides is non-reciprocal. Hence, a non-reciprocal coupling photonic lattice is formed~\cite{Rechtsman2013, PhysRevB.99.201411, Wang_2021, Wang:22}. Besides, the replacement of straight waveguides with staggered helical ones in photonic lattices will result in an additional asymmetric phase factor $ e^{\pm i\varphi} $ in the coupling coefficient, which is also another way to obtain a non-reciprocal coupling photonic lattice~\cite{Li:16, PhysRevLett.117.013902, Maczewsky2017, Ji:18, PhysRevA.100.043808, PhysRevResearch.2.033127, PhysRevB.104.195301, PhysRevB.104.125406, PhysRevResearch.5.L022046}.

Once the intrinsic coupling coefficients of waveguides and the lattice geometry are chosen, the energy band structure is determined, ultimately affecting how light behaves in the lattice. Therefore, designing the band structure is one of the most critical methods to regulate light transmission~\cite{RevModPhys.82.3045, PhysRevApplied.7.014015, PhysRevA.96.043803, PhysRevLett.120.063902, PhysRevA.98.043838, PhysRevA.106.013523, PhysRevA.105.033512, PhysRevA.107.023509}. Since lattice geometry is usually not easily changeable, the lattice intrinsic coupling coefficient naturally becomes the preferred variable for regulating energy bands~\cite{PhysRevA.100.043808, Stojanovic_2022}. Many intriguing phenomena have emerged because of this, such as topological property~\cite{LAN2022100076}, band gap and touch~\cite{PhysRevLett.96.223903, PhysRevB.106.155146}, adjustable flat band~\cite{PhysRevA.100.043808} and band inversion~\cite{PhysRevB.105.035102}.

Flat bands~\cite{GuzmAnSilva2014, PhysRevB.97.045120, PhysRevB.97.035161, PhysRevB.97.195101, PhysRevA.102.023532, PhysRevA.105.L021305}, a type of special energy band structures in condensed physics and photonics, are well known for their zero dispersion in at least one energy band of their spectra. Various flat band lattices have been theoretically and experimentally constructed, such as one-dimensional flat band lattices~\cite{Flach_2014}, Lieb lattices~\cite{PhysRevLett.114.245503}, Kagome lattices~\cite{10.1143/ptp/6.3.306,PhysRevB.98.235109}, and super-honeycomb lattices~\cite{Super-Honeycomb2020}. The most typical common feature of these systems is that their eigenmodes are localized in real space; namely, they all support the existence of compact localized states (CLS)~\cite{PhysRevLett.114.245503, PhysRevB.97.035161, PhysRevB.97.195101, PhysRevA.102.023532, Flach_2014, PhysRevB.98.235109, Super-Honeycomb2020, Daniel2018}. Although various intriguing phenomena have been inspired by the flat band features, such as topological superconductor~\cite{PhysRevB.90.094506}, optical spatial solitons~\cite{Chen_2012}, topological flat band insulator~\cite{PhysRevB.98.245116}, and Anderson localization~\cite{Schwartz2007, PhysRevLett.96.126401}. Most of these studies are performed in reciprocal systems; meanwhile, one structure achieves only a single function. So, realizing flat bands and multifunctional systems in a non-reciprocal system is still an interesting challenge.

In this work, we propose a quasi-one-dimensional non-reciprocal Hermitian hourglass photonic lattice. We find that the Peierls phase term of the non-reciprocal coupling coefficient influences the system's energy band structure and the light energy distribution. By tuning the Peierls phase, we can obtain different energy bands, such as flat band, band inversion, and stable band, which are insensitive to the Peierls phase. We calculate the light energy distribution and observe the energy isolation effect induced by the Peierls phase. We also use Gaussian beams to excite the lattice to confirm the theoretical predictions. Moreover, we show that different excitation conditions can lead to single or double-band excitation. We investigate the CLSs of the flat band and use a novel method to excite the flat band. In addition, we find edge states exist in the hourglass photonic lattice when it is finite. Finally, we introduce a defect in the finite lattice and demonstrate the transition from a single-edge state to a double-edge state.

\section{MODEL AND ANALYSIZE}
\begin{figure}[htp]
	\centering
	\includegraphics[width=\linewidth]{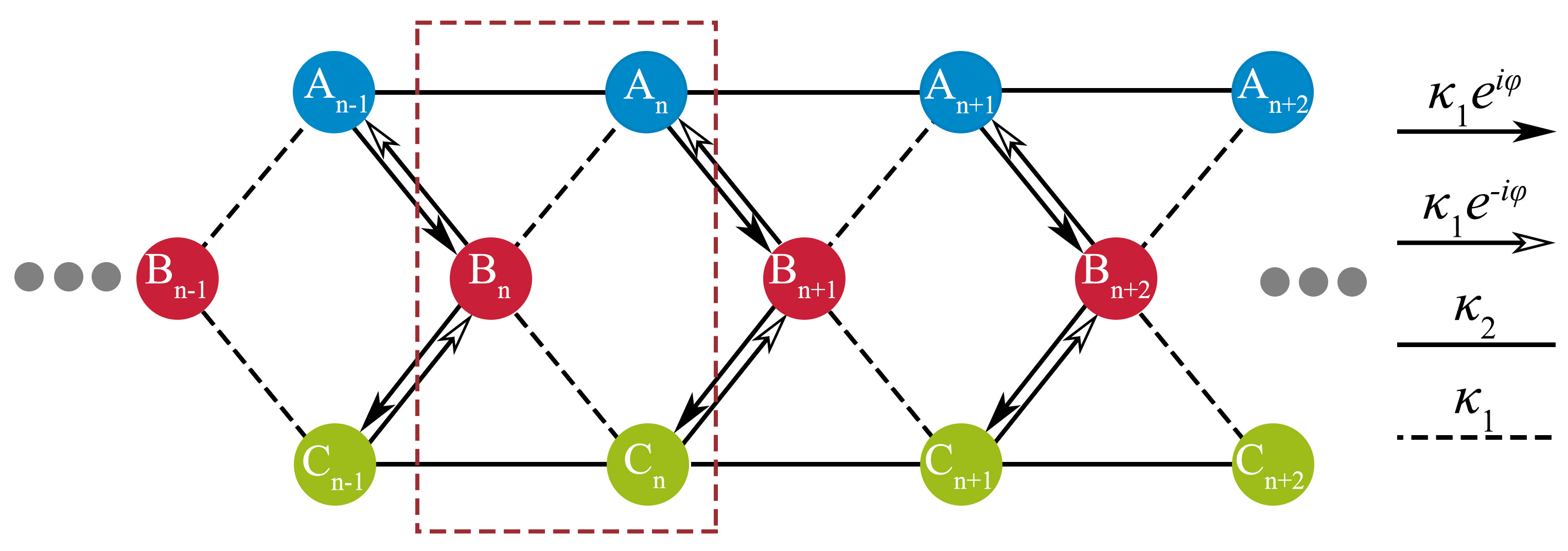}
	\caption{Schematic of the hourglass photonic lattice. The arrows represent the non-reciprocal couplings and photon tunneling direction. The carmine-dashed box represents a unit cell.}
	\label{fig.1}
\end{figure}
We construct an hourglass photonic lattice model that consists of three sites ($\mathrm{A}_n$, $\mathrm{B}_n$, and $\mathrm{C}_n$) per unit cell, as shown in Fig. ~\ref{fig.1}, where $n$ is the lattice site number, $\kappa_1$ is intracell coupling coefficient, $\kappa_2$ is intralayer coupling coefficient, and $\kappa_1e^{\pm i\varphi}$ is intercell coupling coefficient; $\varphi$ is the non-reciprocal coupling phase, also known as the Peierls phase, whose positive or negative sign depends on the direction of photon tunneling. Under periodic boundary conditions, the Hamiltonian of this model in momentum space is given by
\begin{align}
	\begin{aligned}
		& \textbf{\textit{H}}= \\
		& \left(\begin{array}{ccc}
			2 \kappa_2 \cos k & \kappa_1+\kappa_1 e^{i(k-\varphi)} & 0 \\
			\kappa_1+\kappa_1 e^{i(\varphi-k)} & 0 & \kappa_1+\kappa_1 e^{-i(k+\varphi)} \\
			0 & \kappa_1+\kappa_1 e^{i(k+\varphi)} & 2 \kappa_2 \cos k
		\end{array}\right),
	\end{aligned}
	\label{eq:1}
\end{align}
where $k$ is the Bloch wavevector in the first Brillouin zone. By analyzing the Bloch Hamiltonian, we find that it satisfies $\textit{\textbf{H}}=\textit{\textbf{H}}^{\dagger}$ despite the existence of a non-reciprocal coupling coefficient in the lattice, meaning that the system is a Hermitian one with real energy spectra. Solving the Hamiltonian $\textit{\textbf{H}}$, we can obtain the expression of the band structures as follows:
\begin{align}
	\begin{aligned}
		E(k,\varphi)&= 2 \kappa_2 \cos k ,\kappa_2 \cos k \\
		& \pm \frac{\sqrt{2}}{2} \sqrt{8 \kappa_1^2+\kappa_2^2+\kappa_2^2 \cos 2k+8 \kappa_1^2 \cos k \cos \varphi} .
		&
	\end{aligned}
	\label{eq:2}
\end{align}

\begin{figure}[tp]
	\centering
	\includegraphics[width=\linewidth]{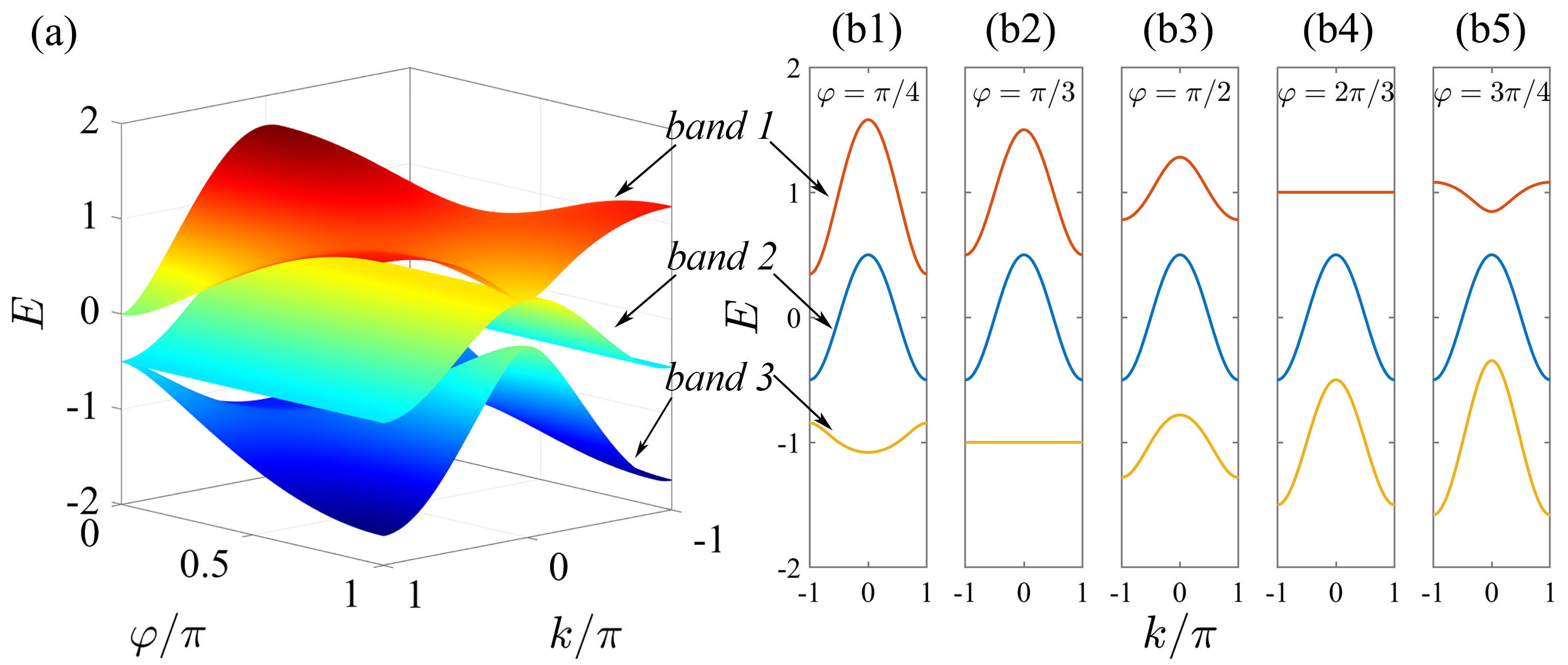}
	\caption{Energy band structures. (a) The surface of bands versus the wavevector $k$ and Peierls phase $\varphi$. (b1)-(b5) The energy band curves under different $\varphi$. Here, $\kappa_1=0.5$, $\kappa_2=0.25$.}
	\label{fig.2}
\end{figure}

According to Eq.~(\ref{eq:2}), we obtain the energy band structure of the system as shown in Fig.~\ref{fig.2}. From Fig.~\ref{fig.2}(a),  we can see that the energy band 2 is stable, while bands 1 and 3 vary with the change of Peierls phase $ \varphi$. The inversion for band 1 is realized by increasing the Peierls phase from $\varphi = \pi/4$ to  $\varphi = 3\pi/4$ [see the orange curves in Fig.~\ref{fig.2}(b1)-(b5)]; similar phenomenon happens with band 3 [see the yellow curves in Fig.~\ref{fig.2}(b1)-(b5)]. Besides, when the Peierls phase $\varphi$ is $\pi/3$ or $2\pi/3$, flat bands appear in the energy band structure, as shown in Figs.~\ref{fig.2}(b2, b4), indicating that the CLSs could exist under these two conditions. 

\begin{figure}[htb]
	\centering
	\includegraphics[width=\linewidth]{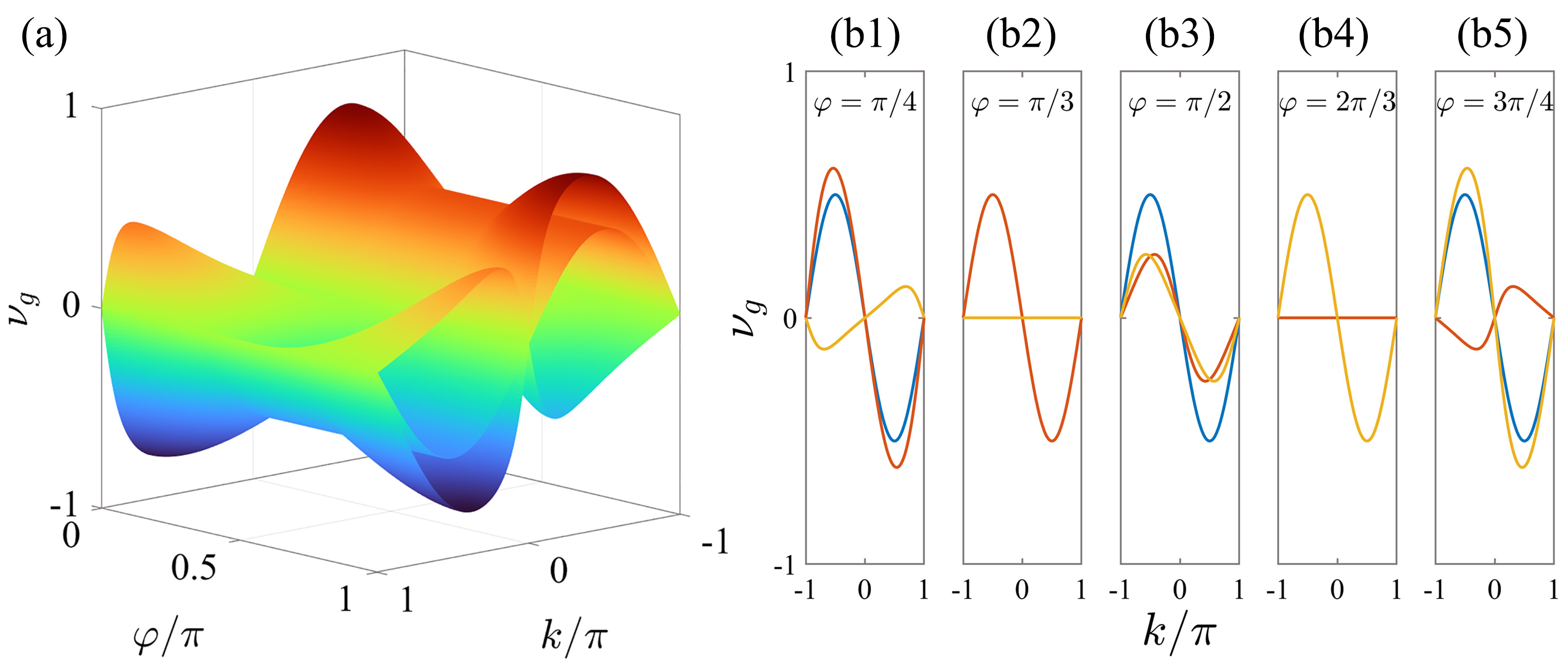}
	\caption{Group velocity surfaces. (a) $v_g$ versus the wavevector $k$ and Peierls phase $\varphi$. (b1)-(b5) $v_g$ under different $\varphi$. Here, $\kappa_1=0.5$, $\kappa_2=0.25$.}
	\label{fig.3}
\end{figure}

We further study the group velocity $v_g$, defining it as $v_g=\partial E/\partial k$. $v_g$ reflects the energy bands' characteristics, usually used to reveal the light propagation behaviors. From Eq.~(\ref{eq:2}), we obtain the expression of group velocity,
\begin{align}
	\begin{aligned}
		v_g=&-2\kappa_2\sin k,-2\kappa_2 \sin k  \\
		&\pm \frac{\sqrt{2}\kappa_2^2 \cos k+2 \kappa_1^2 \cos \varphi\sin k}{\sqrt{8 \kappa_1^2+\kappa_2^2+\kappa_2^2 \cos 2k+8 \kappa_1^2 \cos k \cos\varphi}}.
		&
	\end{aligned}
	\label{eq:3}
\end{align}
We show the surface distribution of $v_g $ with the wave vector$k$ and $\varphi$ in Fig.~\ref{fig.3}(a) using Eq.~(\ref{eq:3}). Figures.~\ref{fig.3}(b2, b3) show that $v_g$ of flat bands are zeroes and the other two $v_g$ curves are degenerate. The blue $v_g $ curve is unchanged with $\varphi$ in Fig.~\ref{fig.3} corresponding to the stable energy band in Fig.~\ref{fig.2}. Moreover, the diffraction-free transmission will be obtained at $(k, \varphi)=(\pm\pi/2, \pi/3)$ or $ (k, \varphi)=(\pm\pi/2, 2\pi/3)$  since the first derivative of $v_g$ is zero [see the extreme points in Figs.~\ref{fig.3}(b2) and (b4)].

\section{ENERGY ISOLATION AND BAND INVERSION}
Based on the above analysis, we conclude that the Peierls phase $\varphi$ governs most of the physical phenomena in the system. Therefore, we theoretically study the impacts of Peierls phase $\varphi$ on the eigenstates of the system with stationary solution method~\cite{Ji:18}. Firstly, we substitute the stationary solution $ (a_n,b_n,c_n)^{T}=\exp(-i\Lambda z)(\mathrm{A}_n,\mathrm{B}_n,\mathrm{C}_n)^T $ into Eq.~(\ref{eq:1}), and obtain the static equation,  
\begin{align}
	\Lambda \zeta_n=\textit{\textbf{H}}\zeta_n,
	\label{eq:4}
\end{align}
where $ \Lambda $ is the stationary frequency, $\zeta_n=(a_n,b_n,c_n)^{T}$ and the corresponding $ (\mathrm{A}_n,\mathrm{B}_n,\mathrm{C}_n)^T $ describes the $ z $-independent amplitudes at the $ n $th waveguide in layers A, B, and C, respectively. Secondly, to simplify the operation, $ (\mathrm{A}_n,\mathrm{B}_n,\mathrm{C}_n)^T $ is expressed as  $ (\mathrm{A}_n,\mathrm{B}_n,\mathrm{C}_n)^T=\{\exp[-i(kn+\delta_a)]$, $\exp[-i(kn+\delta_b)]$, $\exp[-i(kn+\delta_c)]\}(a,b,c)^T $, where $ \delta_a$, $\delta_b $ and $\delta_c $ are the initial phase and set to be zero for simplification and $ (a,b,c)^T $ is $ n $-independent amplitudes in layers A, B, and C. Finally, substituting the assumed solution into Eq.~(\ref{eq:4}), we obtain the energy distribution ratios between the different layers, read as
\begin{gather}
	\left|\frac{\mathrm{A}_n}{\mathrm{~B}_n}\right|^2=\left|\frac{a}{b}\right|^2=\left|\frac{\kappa_1\left[e^{-i(k+\varphi)}+1\right]}{\Lambda-2 \kappa_2 \cos k}\right|^2,\label{eq:5} \\
	\left|\frac{\mathrm{C}_n}{\mathrm{~B}_n}\right|^2=\left|\frac{c}{b}\right|^2=	\left|\frac{\kappa_1\left[e^{-i(k-\varphi)}+1\right]}{\Lambda-2 \kappa_2 \cos k}\right|^2. \label{eq:6}
\end{gather}
Considering the energy conservation of the system, $ |\mathrm{A}_n|^2+|\mathrm{~B}_n|^2+|\mathrm{C}_n|^2=1 $. The results are as follows,
\begin{align}
	\left|\mathrm{A}_n\right|^2 &= \frac{2 \kappa_1^2 [ 1+ \cos (k+\varphi) ]}{f(k, \varphi)} , \label{eq:7}\\ 
	\left|\mathrm{B}_n\right|^2 & = \frac{[\Lambda - 2\kappa_2 \cos k]^2}{f(k, \varphi)} ,         \label{eq:8}\\ 
	\left|\mathrm{C}_n\right|^2 &=   \frac{2 \kappa_1^2 [ 1+ \cos(k-\varphi)] }{f(k, \varphi)},       \label{eq:9}
\end{align}
where Eqs.~(\ref{eq:7}-\ref{eq:9}) hold when $f(k, \varphi) = 4\kappa_1^2 + 2\kappa_2^2 + \Lambda^2 + 2\kappa_2^2 \cos 2k + 4\cos k [\kappa_1^2 \cos\varphi - \kappa_2\Lambda] >0$. We set $\kappa_1=0.5$, $\kappa_2=0.25$, $ \Lambda=1.5 $, $ k\in[-\pi,\pi]$, $ \varphi\in[0,\pi] $, and verify that $f(k, \varphi)>0$ holds for all values of $k$ and $\varphi$.

\begin{figure}[htbp]
	\centering
	\setlength{\belowcaptionskip}{-1pt} 
	\includegraphics[width=\linewidth]{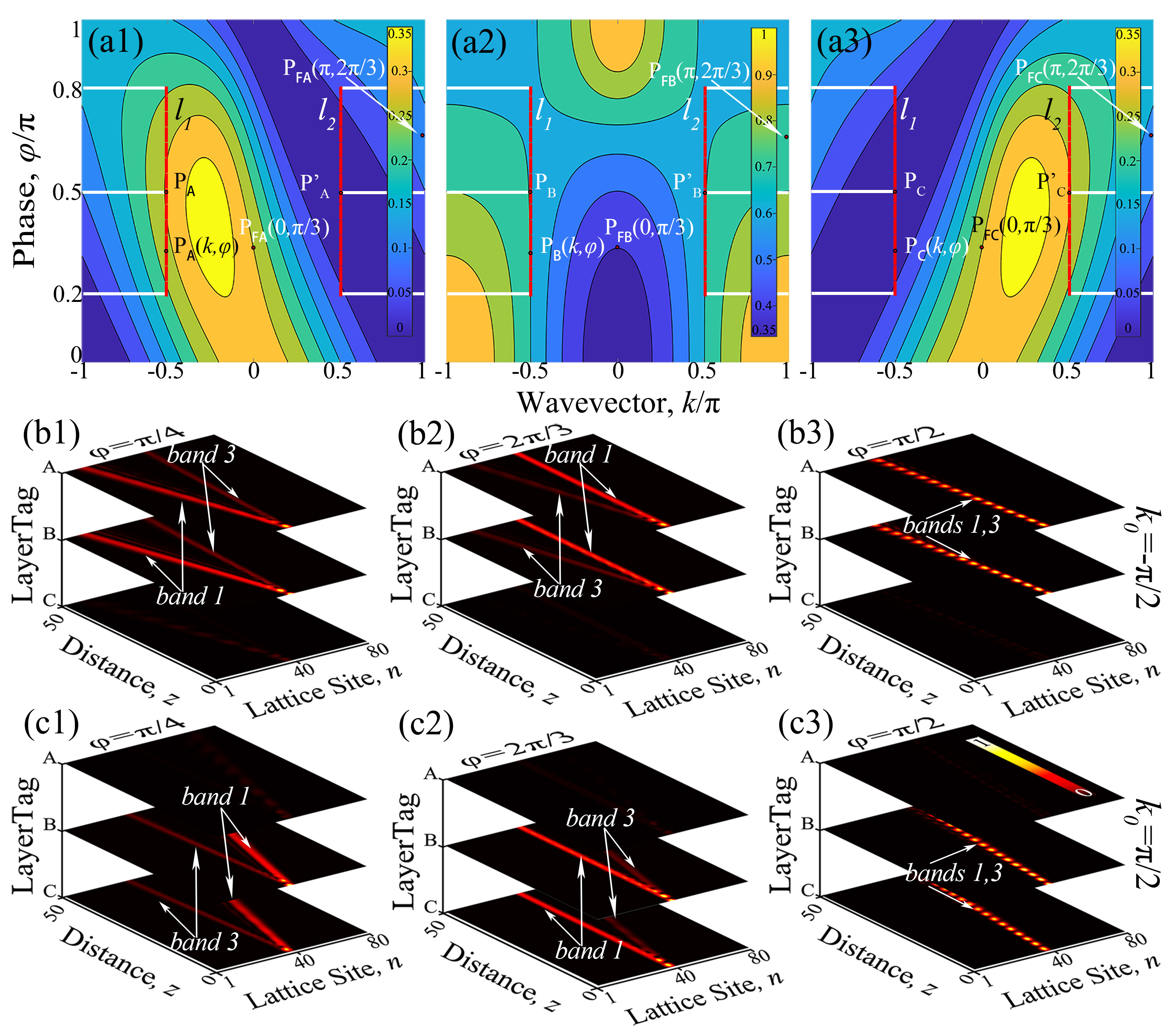}
	\caption{Characterization of light transmission. (a1)-(a3) Maps of light energy distributions of layers A, B, and C versus the Peierls phase $\varphi$ and the wavevector $k$. (b1)-(b3) and (c1)-(c3) Numerically calculated light evolution. Here, $\Lambda$=1.5, $\kappa_1=0.5$, $\kappa_2=0.25$.}
	\label{fig.4}
\end{figure}

According to Eqs.~(\ref{eq:7}-\ref{eq:9}), the light energy distribution maps in three layers with respect to the wavevector $k$ and the Peierls phase $\varphi$ are obtained as shown in Figs.~\ref{fig.4}(a1-a3). When $k=\pm \pi/2$ and $\varphi\in[\pi/5,4\pi/5] $ (see lines $l_1$ and $l_2$), the vast majority of the light energy appears in layers A and B or layers B and C  [Figs.~\ref{fig.4}(a1-a3)]; apparently, these lead to the energy isolation of layer C or layer A, respectively. Further, considering the solutions are stationary, the energy in these three layers will remain invariant once a plane wave is incident at point P($k$, $\varphi$) on lines $l_1$ or $l_2$. 

To verify the above conclusions, we numerically simulate the evolution of three Gaussian beams with intensity functions,
\begin{align}
	\begin{aligned}
		I =\thinspace&U_{A}e^{-(\frac{n_{A}-n_{A0}}{\omega})^2+ik_0n_{A}}\\ &+U_{B}e^{-(\frac{n_{B}-n_{B0}}{\omega})^2+ik_0 n_{B}}\\
		&+U_{C}e^{-(\frac{n_{C}-n_{C0}}{\omega})^2+ik_0n_{C}},
	\end{aligned}
	\label{eq:10}
\end{align}
where $U_{(A, B,C)}$ are the amplitudes, satisfying $U_{A}/U_{B}/U_{C}=I_{A}(k, \varphi)/I_{B}(k, \varphi)/I_{C}(k, \varphi)$, $I_{(A,B,C)}$ are the intensities at points $P_A$, $P_B$ and $P_C$ in Figs.~\ref{fig.4}(a1-a3); $n_{(A,B,C)0}$ are the incident lattice site numbers, $\omega$ is the width of the Gaussian beam and $k_0$ is the initial wavevector.

When $I_{(A, B, C)}$ are at points $P_{(A, B, C)}$ with wavevector $k_0 = -\pi/2$ [see Figs.~\ref{fig.4}(b1-b3)], most of the energy distributes in layers A and B, while the layer C is isolated; When $I_{(A, B, C)}$ are at points $P_{(A^{\prime}, B^{\prime}, C^{\prime})}$ with wavevector $k_0 = \pi/2$ [see Figs.~\ref{fig.4}(c1-c3)], most of the energy distributes in layers B and C, while the layer A is cut off. Besides,  the intensity distribution in each layer depends on the value of Peierls phase $\varphi$. When $\varphi = \pi/4$ or $2\pi/3$, light bifurcates into two sidelobes, respectively [see Figs.~\ref{fig.4}(b1, b2) and \ref{fig.4}(c1, c2)]; In particular, for $\varphi = \pi/2$, oscillating localizations of light beams are supported between lattice layers A and B with incident wavevector $k_0 = -\pi/2$ or B and C with incident wavevector $k_0 = \pi/2$, both propagating along one direction [see Figs.~\ref{fig.4}(b3, c3)]. It should be noted that the ratios among the three layers are slightly deviated from the theoretical values calculated by Eqs.~(\ref{eq:5}) and (\ref{eq:6}), which can be attributed to the use of the Gaussian beams instead of plane waves. Notably, all these propagations are along the perpendicular directions of the tangent of bands 1 and 3. Moreover, two sidelobes in Figs.~\ref{fig.4}(b1, b2) and \ref{fig.4}(c1, c2) exchange their energy band numbers due to the band inversion, which is consistent with the results in Figs.~\ref{fig.2}(b1, b4). These results indicate that the energy band excitation and engineering could be achieved by tuning the parameters of $k_0$ and $\varphi$. 

\section{EXCITATION OF A SINGLE ENERGY BAND AND DOUBLE ENERGY BANDS}
\begin{figure}[htb]
	\centering
	\includegraphics[width=\linewidth]{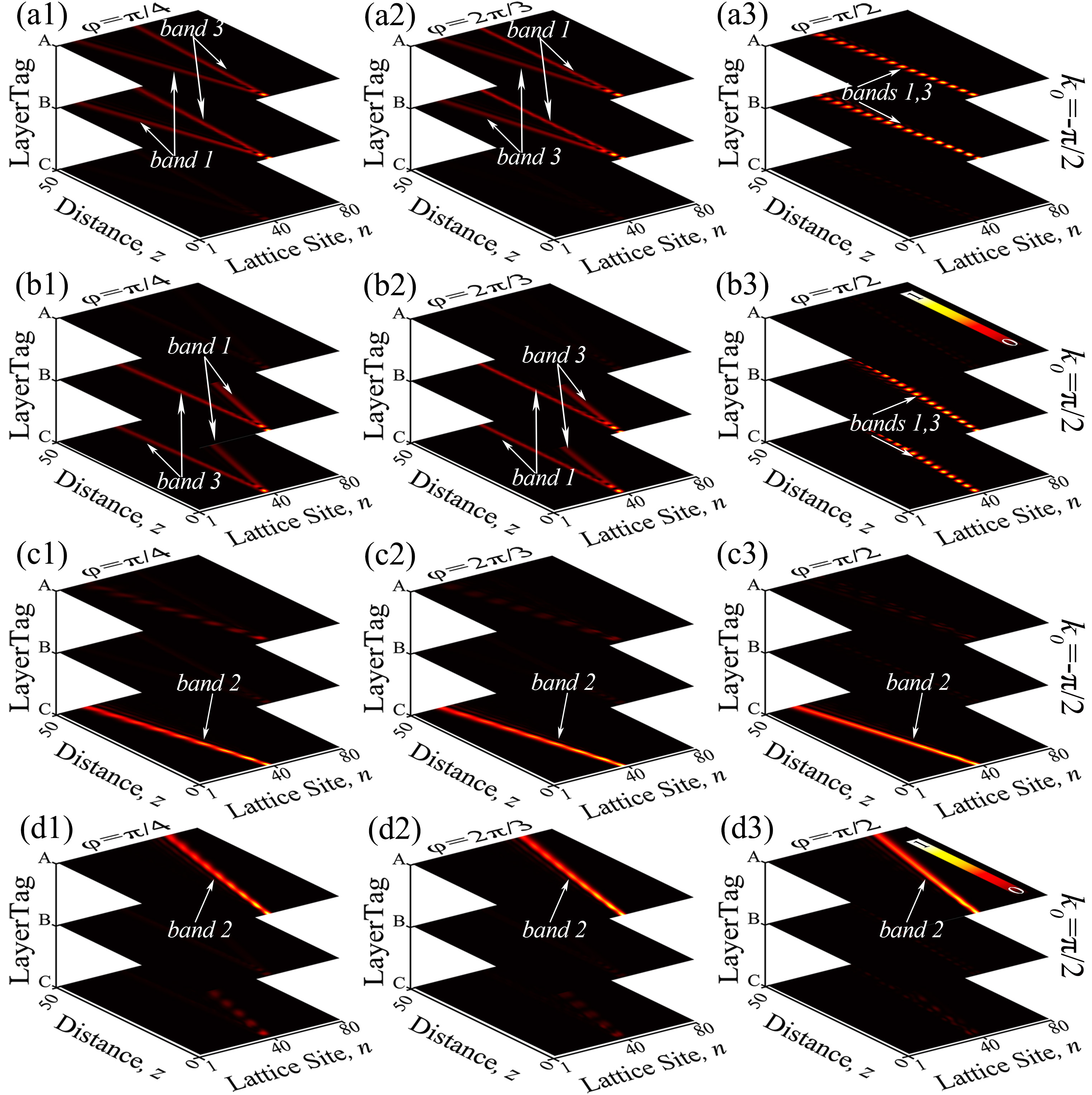}
	\caption{Simulations of single and double energy bands excitations. (a1)-(a3) and (b1)-(b3) Excitation at $k=\mp\pi/2$ in layer B with different $ \varphi $. (c1)-(c3) and (d1)-(d3) Excitation at $ k=\mp\pi/2 $ in layer C and A with different $ \varphi $. Here, $\kappa_1=0.5$, $\kappa_2=0.25$, $\omega=2$.}
	\label{fig.5}
\end{figure}

Based on the conclusions of Section \Rmnum{3}, this section investigates the light transmission properties resulting from a single Gaussian beam incident in the lattice. Figures~\ref{fig.5}(a1-a3) and (b1-b3) illustrate the light wave evolution results when a single Gaussian beam is injected in layer B. We refer to this incident beam as a nearly steady state beam, which satisfies $U_{A}/U_{B}/U_{C}=0/1/0$. From Figs.~\ref{fig.5}(a1-a3) and (b1-b3), energy isolation appears: most energy distributes in layers A and B at $k_0 = -\pi/2$ or layers B and C at $k_0 = \pi/2$. Besides, all transmissions result from the excitation of energy bands 1 and 3 [see Figs.~\ref{fig.5}(a1-a3) and (b1-b3)]. Also, the switch of energy band numbers between two sidelobes due to band inversion exists [see Figs.~\ref{fig.5}(a1, a2) and \ref{fig.5}(b1, b2)]. Additionally, the propagations of the special case that $ \varphi=\pi/2 $ are still oscillatory and localized [see Figs.~\ref{fig.5} (a3, b3)]. These results are all consistent with the conclusion of Section \Rmnum{3} using a steady state beam. 
 However, the main difference is that the energy in sidelobes is no longer mostly concentrated on the excited sidelobe of band 1. The energy in the excited sidelobe of band 3 has become comparable to that of band 1 [see Figs.~\ref{fig.4}(b1, b2, c1, c2) and Figs.~\ref{fig.5} (a1, a2, b1, b2)].

In addition, the excitation of double energy bands involves flat bands [see Figs.~\ref{fig.4}(b2, c2) and Figs.~\ref{fig.5}(a2, b2)]. Because the flat band exists when $ \varphi=2\pi/3 $ [see Figs.~\ref{fig.2}(b2-b4)], and are located in band 1, respectively.

When a single Gaussian beam is input in layer C or A at $k_0 = -\pi/2$ and $k_0 = \pi/2$, respectively, light energy is almost confined in the incident layer C or A due to the energy isolation as shown in Figs.~\ref{fig.5}(c1-c3, d1-d3). The incident beam is called away steady state beam, which has $U_{A}/U_{B}/U_{C}=0/0/1$ at $k_0 = -\pi/2$ and $U_{A}/U_{B}/U_{C}=1/0/0$ at $k_0 = \pi/2$. In contrast to incident steady or nearly steady beam, the propagation does not bifurcate. Also, all lights propagate along the perpendicular direction of the tangent of the stable band 2. Based on the analyses above, the excitation conditions depend not only on the wavevector $k$ and Peierls phase $\varphi$ but also on the initial incident site. Therefore, we can determine a particular excitation band by specifying the values of $\varphi$, $k$, and the incident position.
\section{EXCITATION OF FLAT BANDS}
In Section \Rmnum{4}, we have realized the excitation of the mixed energy band, in which the other curve band still influences the flat band excitation. The CLS mode excitation with multibeam incidence has been studied theoretically and experimentally~\cite{Flach_2014, PhysRevLett.114.245503}. However, the study on flat band excitation with a single beam still needs to be explored. Considering the simple optical path configuration and potential applications in light manipulation, a single beam excitation might be more meaningful, although the physical mechanism is still unclear. 

From Section \Rmnum{1}, the hourglass lattice has two kinds of flat bands, as shown in Figs.~\ref{fig.2}(b2, b4). For $ \kappa_1=0.5 $, $ \kappa_2=0.25 $ and $ \varphi=\pi/3 $, the flat band energy is 
\begin{align}
	E_{FB}=&-1.
\end{align}
When $ \varphi=2\pi/3 $, the flat band energy is
\begin{align}
	E_{FB}=&+1.
\end{align}
\begin{figure}[htp]
	\centering
	\includegraphics[width=\linewidth]{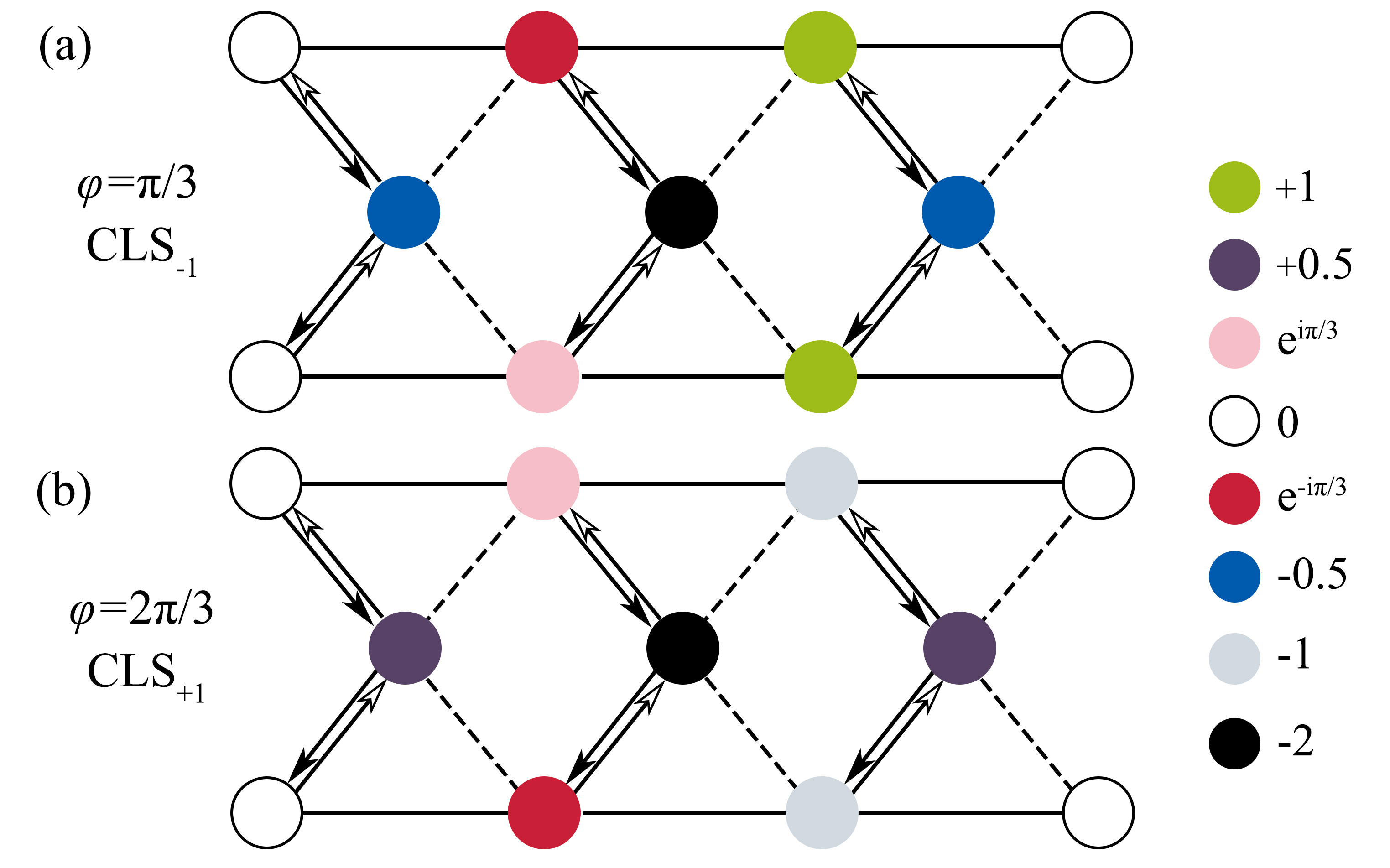}
	\caption{Compact localized states (CLS). (a), (b) CLS mode for $E=-1$ and $E=1$, respectively. The colors represent the filled amplitude.}
	\label{fig.6}
\end{figure}We apply the fundamentals to excite the flat band first and calculate the distribution of wave-function amplitudes at different lattice sites, which can be efficiently evaluated using a Schr\"odinger equation of the form
\begin{align}
	(E-\epsilon_i) \psi_i=\sum_j \kappa_{i j} \psi_j,
	\label{eq:12}
\end{align}
where $ \epsilon_i $ is the on-site potential at the $ i $th site, $ \kappa_{ij} $ is the coupling amplitude between neighboring sites, and $  \psi_i $ is the wave function amplitude at the $ i $th site. Figures~\ref{fig.6}(a, b) illustrate the wave-function amplitude distribution corresponding to the CLSs with energies $ E_{FB} = -1 $ and $ E_{FB} = +1 $. The wave function corresponding to the CLSs is localized over a finite number of lattice sites with non-zero amplitudes (marked by solid-colored circles). Beyond those sites, the amplitude of the wave function is zero (marked by open circles).

\begin{figure}[htp]
	\centering
	\includegraphics[width=\linewidth]{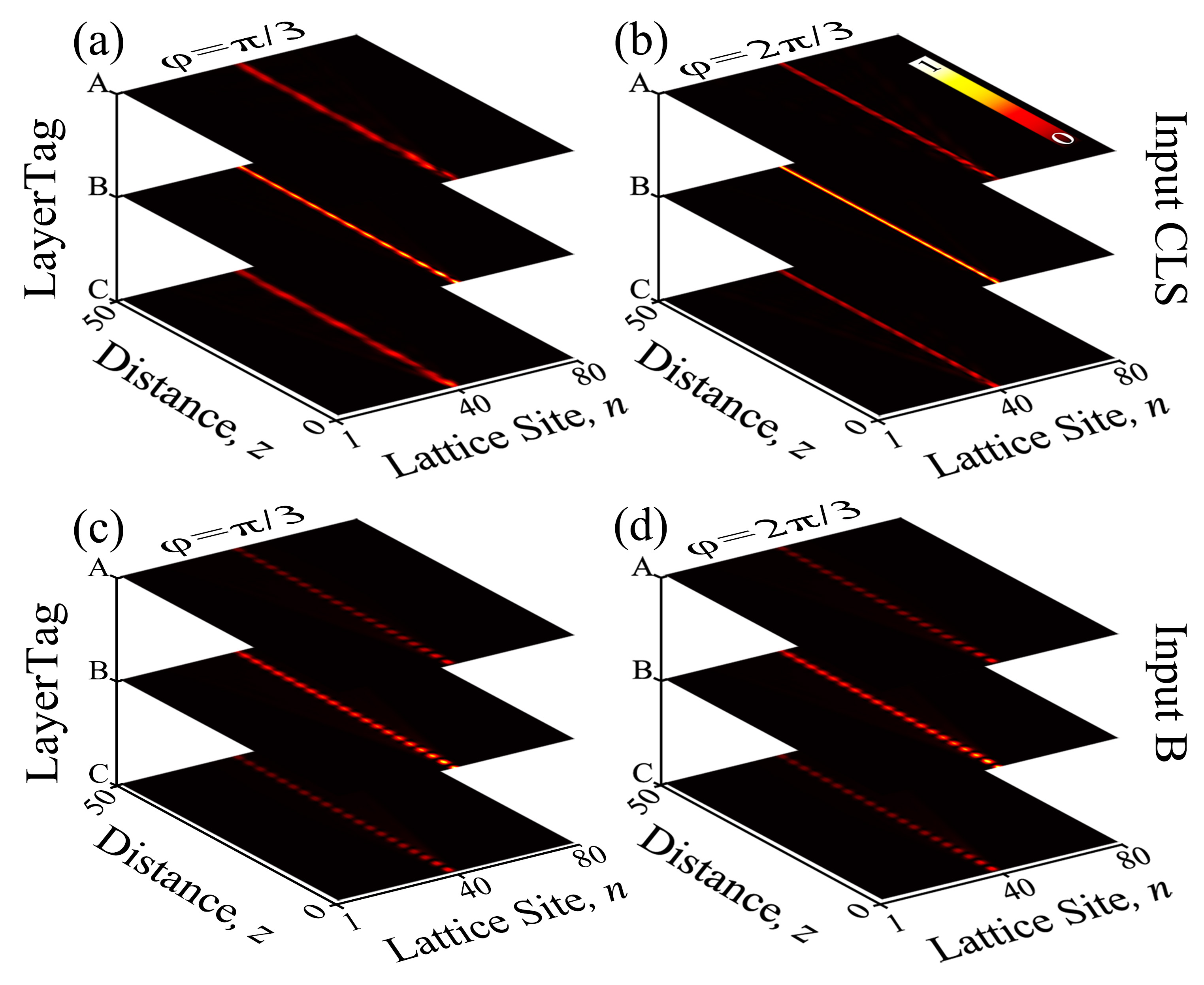}
	\caption{Flat band excitation. (a), (b) Excitations with CLS$_{-1} $ and CLS$_{+1} $ modes at $ k_0= \pi $ when $\omega=1$. (c), (d) Excitation with a single Gaussian beam from layer B at $(k_0, \varphi) = (0, \pi/3)$ and $(k_0, \varphi) = (\pi, 2\pi/3)$, respectively, when $\omega=2$. Here, $\kappa_1=0.5$, $\kappa_2=0.25$.}
	\label{fig.7}
\end{figure}

Figures~\ref{fig.7}(a, b) show the numerical simulation of light propagation by inputting  Gaussian beam arrays, which constitute a CLS mode. All the light beams propagate straightly and locally in layers A, B, and C, and the energy primarily concentrates in layer B, demonstrating that the CLS mode in Fig.~\ref{fig.6} is the flat band mode exactly. Though this method successfully excites a single flat band, it is complex. 

In the following, we further study the operation principle of a single flat band excitation with a single beam incidence. We employ a single Gaussian beam to excite a single flat band. We inject the beam with normal incidence ($ k_0=0 $) into layer B with $\varphi=\pi/3$. As seen in Fig.~\ref{fig.7}(c), light propagates straightly and locally in three layers, showing that a simple excitation method is practicable. Similarly, light also achieves the diffraction-free oscillating localization among the three layers when a Gaussian beam with wavevector $k_0=\pi$ is incident into layer B with $\varphi=2\pi/3$ [Fig.~\ref{fig.7}(d)]. Besides, most of the light energy in Figs.~\ref{fig.7}(c, d) is in layer B, and the rest is equally distributed in layers A and C. The reason for these phenomena can be found in Figs.~\ref{fig.4}(a1-a3), the intensities in waveguides $\mathrm{A}_n$, $\mathrm{B}_n$ and $\mathrm{C}_n$ satisfy $|\mathrm{A}_n|^2=|\mathrm{C}_n|^2<|\mathrm{B}_n|^2$ at $ k=0$, $\pi$ when $\varphi=\pi/3$, $2\pi/3$, respectively [attention to the intensity value represented by the color in points $ \mathrm{P}_{\mathrm{FA(B, C)}}(0,\pi/3) $ and $ \mathrm{P}_{\mathrm{FA(B, C)}}(\pi,2\pi/3) $]. In addition, the oscillating localization among the three layers can only be caused by the single flat band excitation since the group velocity dispersions for the other two bands are non-zeroes. Our method provides a simple way to achieve individual flat band excitation and provides a novel perspective on energy band manipulation.

\section{EDGE STATES}
\begin{figure}[htp]
	\centering
	\includegraphics[width=\linewidth]{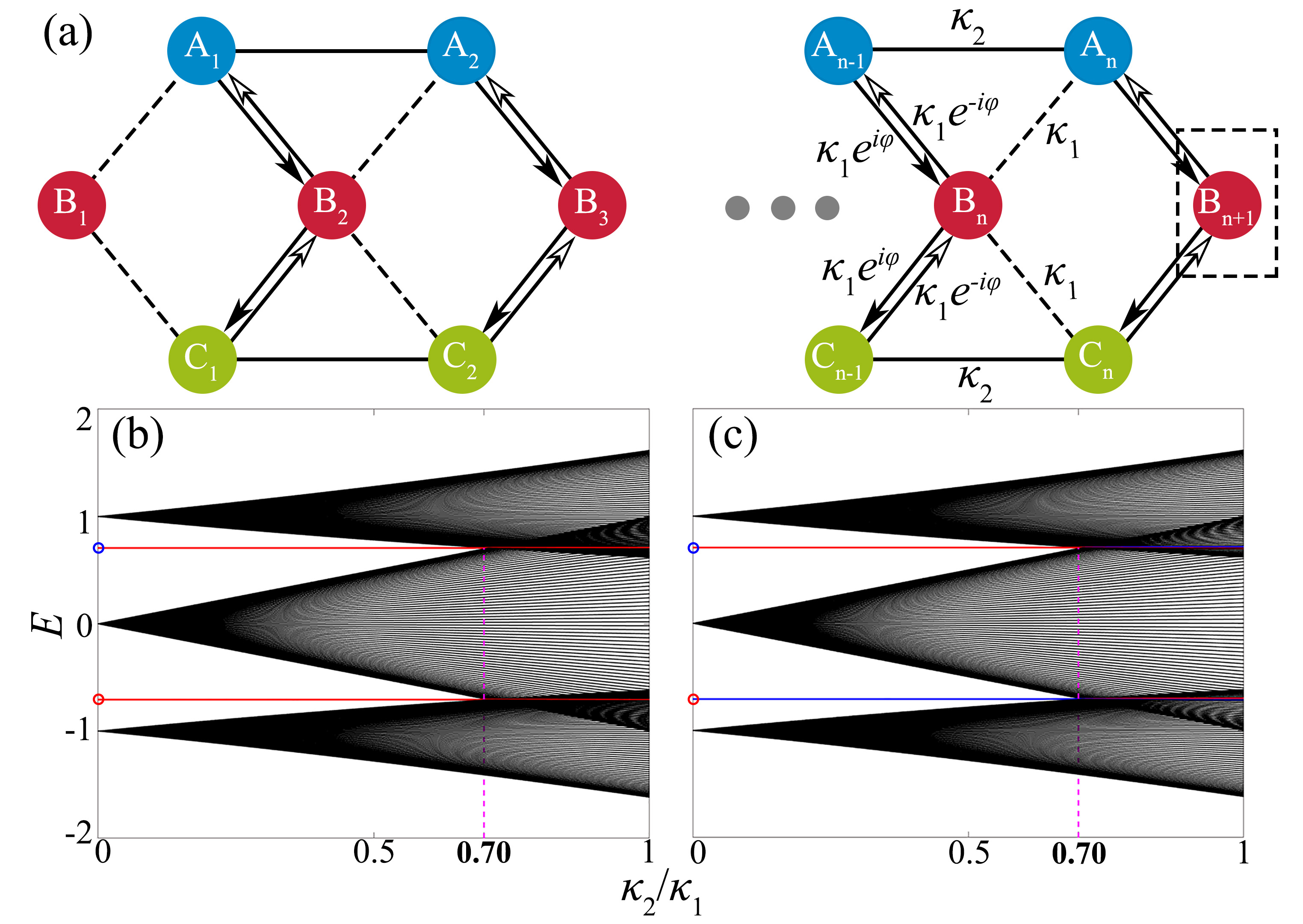}
	\caption{(a) Schematic of the hourglass photonic lattice under an open boundary condition. (b) Energy spectra without the defect site $\rm B_{n+1} $ under open boundary conditions. (c) Energy spectra with the defect site $\rm B_{n+1} $ under open boundary conditions. Here, $\kappa_1=0.5,\varphi=\pi/2,n=80$, blue circle position $ E = \sqrt{3} $, red circle position $ E = -\sqrt{3} $.}
	\label{fig.8}
\end{figure}
After studying the bulk states, in this section, we will introduce the edge states in the hourglass photonic lattice, including single-edge and double-edge states. As shown in Fig.~\ref{fig.8}(a), the lattice is under an open boundary condition, and the site $ \rm B_{n+1} $ is considered a defect. When site $\rm B_{n+1} $ is absent, the energy spectra are shown in Fig.~\ref{fig.8}(b). It can be seen from Fig~\ref{fig.8}(b) that when $\varphi=\pi/2$ and $\kappa_2/\kappa_1<0.70$, there are two non-zero edge states, while when $\varphi=\pi/2$ and $\kappa_2/\kappa_1<0.70$, these two edge states disappear. After adding site $\rm B_{n+1} $, we obtain the energy spectrum as shown in Fig.~\ref{fig.8}(c). When $\varphi=\pi/2$ and $\kappa_2/\kappa_1<0.70$, there are four non-zero edge states, but they degenerate in pairs, meaning that the edge states undergo double degeneration, and the number of edge states changes from two to four.

\begin{figure}[htp]
	\centering
	\includegraphics[width=\linewidth]{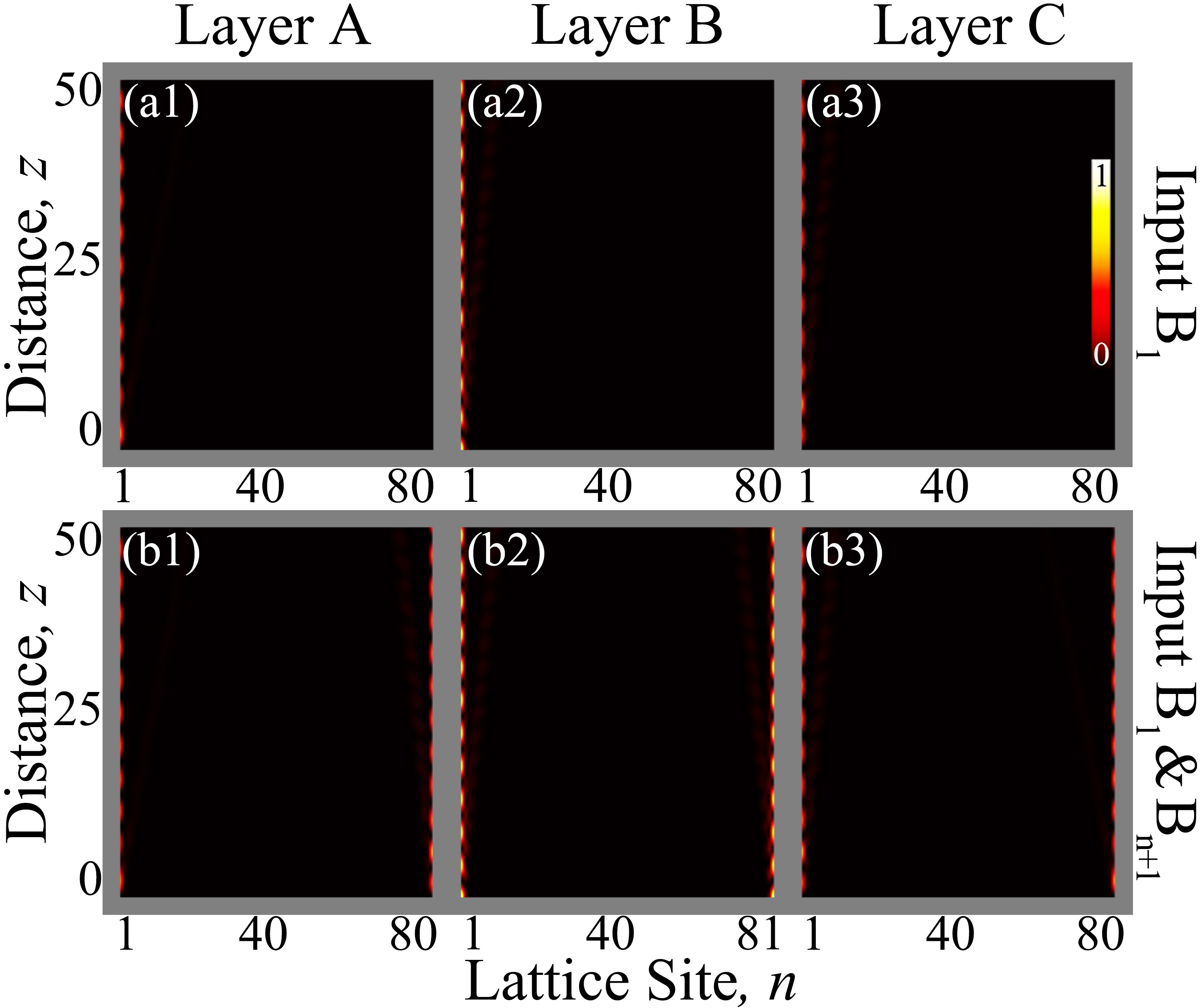}
	\caption{Simulation of light propagation on the lattice edge. (a1)-(a3) Single edge state. (b1)-(b3) Double edge state. Here, $ n=80 $ $\kappa_{1}=0.5$, $\omega=1$.}
	\label{fig.9}
\end{figure}

For the edge states, the closer they are to the bulk energy spectra, the more affected they are. Therefore, we choose the ratio of $\kappa_2/\kappa_1=0.4$ to simulate the light propagation. Figs.~\ref{fig.9}(a1-a3) shows the numerical simulations of the propagation behavior for a single Gaussian beam incident at waveguide $\rm B_{1}$. Light only oscillates among waveguides $\rm A_{1}$, $\rm B_{1}$ and $\rm C_{1}$ without obvious diffraction to adjacent waveguides. We attribute such an oscillating localization to the interference of single-side edge states in the lattice with defect-free edge [see Figs.~\ref{fig.8}(a) and (b)]. Once a defect point $\rm B_{n+1}$ is added, double-side oscillating localizations occur in all three layers when we inject two Gaussian beams at $\rm B_{1} $ and  $\rm B_{n+1} $, respectively [as shown in Figs.~\ref{fig.9}(b1-b3)], indicating that such a system supports edge states on both sides. The oscillating localizations on the left edges remain unchanged compared with those in Figs.~\ref{fig.9}(a1-a3); meanwhile, their energy is almost equal to that of the right edge states. That is to say, the defect transforms the system edge states from single-edge states to double-edge states.

\section{Conclusions}
We have investigated a quasi-one-dimensional non-reciprocal Hermitian hourglass photonic lattice in which energy bands are tunable by changing the non-reciprocal coupling coefficient $ \varphi $. The lattice can realize energy band inversion and flat bands as $ \varphi $ changes. By calculating the interlayer distribution of light energy, we found that the system has an energy isolation effect: layers A and B are isolated from layer C when $ k=-\pi/2 $, or layers B and C are isolated from layer A when $ k=-\pi/2 $. Subsequently, during the verification process, we can determine the specific excitation band once we confirm the values of $\varphi$, $k$, and the incident position. In flat band excitation, we can use CLSs to excite flat bands and a more straightforward single beam to excite flat bands. In addition, after adding a defect with an open boundary condition,  a single-edge state of the lattice becomes a double-edge state. Our research results may provide opportunities to explore new theories of energy band regulation in artificial photonic lattices.
\section{Acknowledgement}
We gratefully acknowledge financial support from the National Natural Science Foundation of China (NSFC)(N0.1217040857).

\nocite{*}
\bibliography{ref}

\end{document}